\documentclass[showpacs,twocolumn]{revtex4}

\usepackage{graphicx}
\usepackage{dcolumn}
\usepackage{amsmath}

\makeatletter
\def\btt#1{\texttt{\@backslashchar#1}}
\DeclareRobustCommand\bblash{\btt{\@backslashchar}} \makeatother

\begin{document}

\title[]{Spinning Higher Dimensional Einstein-Yang-Mills black holes}
\author{Sushant~G.~Ghosh$^{a,\;b\;}$} \email{sghosh2@jmi.ac.in,
sgghosh@gmail.com}
\author{Uma Papnoi$^{a}$} \email{uma.papnoi@gmail.com}
\affiliation{$^{a}$ Centre for Theoretical Physics, Jamia Millia
Islamia, New Delhi 110025, India} \affiliation{$^{b}$ Astrophysics
and Cosmology Research Unit, School of Mathematical Sciences,
University of Kwa-Zulu-Natal, Private Bag 54001, Durban 4000, South
Africa}
\date{\today}

\begin{abstract}
We construct a Kerr-Newman-like spacetimes starting from higher
dimensional (HD) Einstein-Yang-Mills black holes via complex
transformations suggested by Newman-Janis. The new metrics are HD
generalization of Kerr-Newman spacetimes which has a geometry
precisely that of Kerr-Newman in 4D corresponding to Yang-Mills (YM)
gauge charge, but the sign of charge term gets flipped in the HD
spacetimes. It is interesting to note that gravitational
contribution of YM gauge charge, in HD, is indeed opposite
(attractive rather than repulsive) that of Maxwell charge. The
effect of YM gauge charge on the structure and location of static
limit surface and apparent horizon is discussed. We find that static
limit surfaces become less prolate with increase in dimensions and
are also sensitive to YM gauge charge thereby affecting the shape of
ergosphere.  We also analyze some thermodynamical properties of
these BHs.
\end{abstract}


\maketitle

\tolerance=5000
\newpage

\section{Introduction}
The Kerr metric \cite{kerr} is an explicit exact solution of
Einstein field equations describing a spinning black hole (BH) in
four dimensional (4D) spacetime. It is well known that BH with
non-zero spinning parameter, i.e., Kerr BH enjoys many interesting
properties distinct from its non-spinning counterpart, i.e., from
Schwarzschild BH \cite{scht}. However, there is a surprising
connection between the two BHs of Einstein theory, and is analyzed
by Newman and Janis \cite{nja}. They demonstrated that applying a
complex coordinate transformation, it was possible to construct both
the Kerr and Kerr-Newman solutions starting from the Schwarzschild
metric and Reissner-Nordstr$\ddot{o}$m metric respectively
\cite{nja}. The Kerr-Newman describes the exterior of a spinning
massive charged BH \cite{Kn}. The Newman Janis Algorithm (NJA) is
successful in analyzing several spinning BH metric starting from
their non-spinning counterparts
\cite{my,Xu,Drake,Kim,Yazadjiev,Drake1,Qw,Caravelli,Modesto,Capo,Johannsen,Bambi,Glass}.
For  a review on the NJA (see, e.g., \cite{d'Inverno}). However, the
NJA has often be considered that there is arbitrariness  and physics
considered this as an \textit{adhoc} procedure \cite{flaherty}. But
Schiffer et. al. \cite{schiffer} gave a very elegant mathematical
proof as why Kerr metric can be considered as complex transformation
of Schwarzschild metric.

It is rather well established that higher dimensions provide a
natural playground for the string theory and they are also required
for its consistency \cite{sv,er}.  Even from the classical
standpoint, it is interesting to study the higher dimensional (HD)
extension of Einstein's theory, and in particular its BH solutions
\cite{kp}.   There seems to be a general belief that endowing
general relativity with a tunable parameter namely the spacetime
dimension should also lead to valuable insights into the nature of
the theory, in particular into its most basic objects: BHs. For
instance, 4D BHs are known to have a number of remarkable features,
such as uniqueness, spherical topology, dynamical stability, and the
laws of BH mechanics. One would like to know which of these are
peculiar to 4D, and which are true more generally? At the very
least, such probings into HD will lead to a deeper understanding of
classical BHs and of what space-time can do at its most extreme.
There is a growing realization  that the physics of HD BHs can be
markedly different, and much richer than its counterpart in 4D
\cite{scht,rcm,aad,Myers}, e.g.,  the event horizon may not be
spherical in HD and also no BH uniqueness \cite{er}. It is of
interest to consider models based on different interacting fields
including the Yang-Mills (YM). In general, it is difficult to tackle
Einstein-Yang-Mills (EYM) equations because of the non-linearity
both in the gauge fields as well as in the gravitational field. The
solutions of the classical YM fields depend upon the particular
\textit{ansatz} one chooses. Wu and Yang \cite{wy} found static
spherically symmetric solutions of the YM equations in flat space
for the gauge group SO(3). A curved spacetime generalization of
these models has been investigated by several authors (see, e.g.,
\cite{py}). Indeed Yasskin \cite{py} has presented an explicit
procedure based on the Wu-Yang \textit{ansatz} \cite{wy} which gives
the solution of EYM rather trivially. Using this procedure,
Mazharimousavi and Halilsoy \cite{shmh07,shmh08,shmh081} have found
a sequence of static spherically symmetric HD EYM BH solutions. The
remarkable feature of this \textit{ansatz} is that the field has no
contribution from gradient; instead, it has pure YM non-Abelian
component. It, therefore, has only the magnetic part.

The strategy of obtaining the familiar Kerr-Newman solution, both in
4D and HD, in general relativity is based on either using the metric
ansatz in the Kerr-Schild form or applying the method of complex
coordinate transformation to a non-rotating charged black hole.
Surprisingly, it has been demonstrated that when employing to HD
dimensional spacetime both approaches lead to same result
\cite{Aliev,Aliev1} The main purpose of this work is to apply NJA to
HD EYM BH metric previously discovered in
\cite{shmh07,shmh08,shmh081} and spinning HD EYM BH metric is
obtained. This result shows that NJA works well also in HD
spacetime. We further discuss the properties of the spinning HD EYM
BH such as horizons and ergosphere. Spinning BH solutions in higher
dimensions are known as Myers-Perry BHs \cite{Myers}. The
thermodynamical quantities associated with the spinning HD EYM BH
are also calculated.  Further we demonstrate that the
thermodynamical quantities of this BH go over to corresponding
quantities of Myers-Perry BH and Kerr BH.

\section{HIGHER DIMENSIONAL STATIC BLACK HOLE IN EINSTEIN-YANG-MILLS THEORY}
We consider $(N+1)(N+2)/2$ parameter Lie group with structure
constant $C_{\left( \beta\right) \left( \gamma\right) }^{\left(
\alpha \right) }$. The gauge potentials $A_{a }^{\left(
\alpha\right)}$ and the YM fields $F_{a b }^{\left(
\alpha\right) }$ are related through the equation
\begin{equation}
F_{a b }^{\left( \alpha\right) }=\partial _{a }A_{b }^{\left(
\alpha\right) }-\partial _{b }A_{a }^{\left( \alpha\right)
}+\frac{1}{2\sigma }C_{\left( \beta\right) \left( \gamma\right)
}^{\left( \alpha \right) }A_{a }^{\left( \beta\right) }A_{b
}^{\left( \gamma \right) }.
\end{equation}
Then one can choose the gravity and gauge field action
(EYM), which in $(N+3)$-dimensions reads \cite{shmh07,shmh08,Ghosh}:
\begin{equation}
\mathcal{I_{G}}=\frac{1}{2}\int_{{M}}dx^{N+3}\sqrt{-g}\left[
R-\sum_{\alpha=1}^{\left(N+1\right) (N+2)/2} F_{a b }^{(\alpha)}F^{(\alpha)a b }%
\right].
\end{equation}
Here, $g$ = det($g_{ab}$) is the determinant of the metric tensor,
$R$ is the Ricci Scalar and $A_{a }^{\left( \alpha\right)} $
are the gauge potentials. We note that the internal indices $
\{\alpha,\beta,\gamma,...\}$ do not differ whether in covariant or
contravariant form. We introduce the Wu-Yang \textit{ansatz} in
$(N+3)$-dimension \cite{shmh07,shmh08,shmh081} as
\begin{eqnarray}
A^{(\alpha)} &=&\frac{Q}{r^{2}}\left( x_{i}dx_{j}-x_{j}dx_{i}\right) \\
2 &\leq &i\leq N+2,  \notag \\
1 &\leq &j\leq i-1 , \notag \\
1 &\leq &\left( \alpha\right) \leq \left( N+1\right) (N+2)/2,\notag
\end{eqnarray}
where the super indices $\alpha$ is chosen according to the values
of $i$ and $j$ in order and we choose $\sigma = Q$
\cite{shmh07,shmh08,shmh081}. The Wu-Yang solution appears highly
non-linear because of mixing between spacetime indices and gauge
group indices. However, it is linear as expressed in the non-linear
gauge fields because purely magnetic gauge charge is chosen along
with position dependent gauge field transformation \cite{py}.
The YM field 2 form is defined by the expression%
\begin{equation}
F^{\left( \alpha\right) }=dA^{\left( \alpha\right)
}+\frac{1}{2Q}C_{\left( \beta\right) \left( \gamma\right) }^{\left(
\alpha\right) }A^{\left( \beta\right) }\wedge A^{\left(
\gamma\right) }.
\end{equation}
The integrability conditions
\begin{equation}
dF^{\left( \alpha\right) }+\frac{1}{Q}C_{\left( \beta\right) \left(
c\right) }^{\left( \alpha\right) }A^{\left( \beta\right) }\wedge
F^{\left( \gamma\right) }=0,
\end{equation}
as well as the YM equations
\begin{equation}
d\ast F^{\left( \alpha\right) }+\frac{1}{Q}C_{\left( \beta\right)
\left( \gamma\right) }^{\left( \alpha\right) }A^{\left( \beta\right)
}\wedge \ast F^{\left( \gamma\right) }=0,
\end{equation}
are all satisfied. Here $d$ is exterior derivative, $\wedge$ stands
for wedge product and $\ast$ represents Hodge duality. All these are in the usual exterior differential forms notation.

Variation of the action with respect to the space-time metric $g_{ab}$
yields the EYM equations
\begin{equation}\label{eq:ee}
G_{a b }=T_{a b}.
\end{equation}
where the gauge stress-energy tensor is
\begin{eqnarray}\label{emt}
T_{a b }=\sum_{\alpha=1}^{\left(N+1\right) (N+2)/2} \left[ 2F_{a
}^{\left( \alpha\right) \lambda }F_{b \lambda }^{\left(
\alpha\right) }-\frac{1}{2}F_{\lambda \sigma }^{\left( \alpha\right)
}F^{\left( \alpha\right) \lambda \sigma }g_{a b }\right].
\end{eqnarray} In general, it is difficult to solve EYM Eq. (\ref{eq:ee}). However, the Wu-Yang \textit{ansatz} \cite{wy} facilitate in obtaining the solution.

The metric for the HD EYM BH \cite{shmh081} obtained using Wu-Yang \textit{ansatz} \cite{wy} is given by
\begin{equation}
ds^2 = f(r)\;  dt^2 - f(r)^{-1}\; dr^2 - r^2 d \Omega_{N+1}^2,
\label{eq:me2}
\end{equation}with
\begin{eqnarray*}
f(r) &=& 1 - \frac{\mu}{r^{N}}- \frac{N}{(N-2)} \frac{Q^2}{r^2},
\;\; N\neq 2,
\end{eqnarray*}
where
\begin{eqnarray*}
d \Omega_{N+1}^2 & = & d \theta^2_{1} + \sin^2{\theta}_1 d
\theta^2_{2} + \sin^2{\theta}_1 \sin^2{\theta}_2d \theta^2_{3} +
\ldots \\  & & + \left[\left( \prod_{j=1}^{N} \sin^2{\theta}_j
\right) d \theta^2_{N+1} \right],
\end{eqnarray*}
where, $\mu$ is the integration constant which can be related to
mass $M$ and $D=N+3$ is spacetime dimensions. Since $T_{ab}$ go as
$r^{-4}$ (same as for Maxwell field in 4D),
interestingly for all $D\geq6$. That is why its contribution in $f(r)$
is same for all $D\geq6$ as in Reissner-Nordstr$\ddot{o}$m.
There is, however, an important difference in the sign before
$Q^2/r^2$ term. In 4D case, it is exactly like Reissner-Nordstr$\ddot{o}$m BH, i.e.,
positive. In contrast, to 4D, the sign before $Q^2/r^2$ is
negative for $D\geq6$. On the other hand, if YM gauge charge is switched off $(Q=0)$, the metric (\ref{eq:me2}) reduces to well known Schwarzschild-Tanghelini metric \cite{scht}. In addition if $N=1$, one may note that it reduces to Schwarzschild metric. When $N=1$, the metric (\ref{eq:me2}) is exactly Reissner-Nordstr$\ddot{o}$m BH with $Q$ as YM gauge charge.
\section{SPINNING HD EYM BH VIA NEWMAN-JANIS ALGORITHM}
We want to derive axially symmetric spinning analogue of the
static  spherically symmetric EYM BH adapting Newman-Janis
\cite{nja} complex transformation. Newman et
al. \cite{nja} discovered curious derivations of stationary,
spinning metric solutions from static, spherically symmetric
solutions in 4D Einstein theory. In order to derive spinning HD EYM BH, we start with the non-spinning version of the HD EYM BH metric (\ref{eq:me2}), with $f(r)$ given by
\begin{eqnarray*}
f(r) &=& 1- \frac{\mu}{r^{N}} - \frac{Q_0^2}{r^2}, \;\; Q_0^2= \frac{N}{(N-2)}Q^2,
\end{eqnarray*}
as a seed solution to construct its spinning counterpart. Following
Newman and Janis \cite{nja}  the first step is to write the metric
(\ref{eq:me2}) in advanced Eddington-Finkelstein coordinates by the
following coordinate transformation:
\begin{eqnarray}
du &=& dt- f(r)^{-1} dr, \label{eq:du}
\end{eqnarray}
we obtain
\begin{eqnarray}
ds^2 = f(r) du^2+ 2 du dr  -r^2d\Omega_{N+1}^2. \label{eq:me3}
\end{eqnarray}
The metric (\ref{eq:me3}) can be
written in terms of a null veiltrad $ Z^a =
(l^a,\;n^a,\;m_{1}^a,\;\bar{m_{1}}^a,\;m_{2}^a,\;\bar{m_{2}}^a,\; \ldots, \;
m_{(N+1)/2}^a,\;   \bar{m}_{(N+1)/2}^b), $ \cite{nja, Capo} as
\begin{eqnarray}
{g}^{ab} = l^a n^b + l^b n^a - m_{1}^a \bar{m}_{1}^b - m_{2}^a
\bar{m}_{2}^b,\;\ldots,\;\\ \nonumber -m_{(N+1)/2}^a  \bar{m}_{(N+1)/2}^b,
\label{NPmetric}
\end{eqnarray} where null vieltrad are \begin{eqnarray*}
l^a &=& \delta^a_r,\\
n^a &=& \left[ \delta^a_u - \frac{1}{2} \left(1- \frac{\mu}{r^{N}}- \frac{Q_0^2}{r^2}\right) \delta^a_r \right],\\
 m_{k}^a &=& \frac{1}{\sqrt{2}r\sin\theta_{1}\sin\theta_{2}, \;\ldots, \sin\theta_{(k-1)}}  \left( \delta^a_{\theta_{k}}
  + \frac{i}{\sin\theta_{k}} \delta^a_{\theta_{(k+1)}} \right),\\
\end{eqnarray*}
where $D=N+3$, is spacetime dimension and $k$ is used to denote the number of vector which take values $1, 2, 3, \;\ldots,\; (N+1)/2$, e.g., in 6D $k$ is 2 which corresponds to following vector
\begin{eqnarray*}
m_1^a &=& \frac{1}{\sqrt{2}r} \left (\delta^a_{\theta_1} + \frac{i}{\sin\theta_{1}} \delta^a_{\theta_2} \right), \nonumber \\
m_2^a &=& \frac{1}{\sqrt{2}r \sin\theta_1} \left (\delta^a_{\theta_2} + \frac{i}{\sin\theta_{2}} \delta^a_{\theta_3} \right).
\end{eqnarray*}
Here we assume $N$ is odd, i.e., spacetime dimension $D$ is even.
However, the final result is independent of this assumption.  Note
$l^a$ and $n^a$ are real, $m_k^a$, $\bar{m_k}^a$ are mutual complex
conjugate. This vieltrad is orthonormal and obeying metric
conditions.
\begin{eqnarray}
l_{a}l^{a} = n_{a}n^{a} = ({m_k})_{a} ({m_k})^{a} = (\bar{m_k})_{a} (\bar{m_k})^{a}= 0,  \nonumber \\
l_{a}({m_i})^{a} = l_{a}(\bar{m_k})^{a} = n_{a}({m_k})^{a} = n_{a}(\bar{m_k})^{a}= 0, \; \nonumber \\
l_a n^a = 1, \; ({m_k})_{a} (\bar{m_k})^{a} = 1. \nonumber
\end{eqnarray}
Now we allow for some $r$ factor in the null vectors to take on complex values. Following \cite{my,Qw}, we rewrite the null vectors in the form
\begin{eqnarray*}
l^a &=& \delta^a_r, \\
n^a &=& \left[ \delta^a_u - \frac{1}{2} \left(1- \frac{\mu}{2r^{N-1}}\left[\frac{1}{r}+\frac{1}{\bar{r}}\right]- \frac{Q_0^2}{r \bar{r}}\right) \delta^a_r \right], \\
m_{k}^a &=& \frac{1}{\sqrt{2}\bar{r}\sin\theta_{1}\sin\theta_{2}\; \ldots \;\sin\theta_{(k-1)}}  \left( \delta^a_{\theta_{k}} + \frac{i}{\sin\theta_{k}} \delta^a_{\theta_{(k+1)}} \right),\\
\end{eqnarray*}
with $\bar{r}$ being the complex conjugate of $r$. In 4D there is
only one possible spinning axisymmetric  spacetime,  and there is
therefore only one angular momentum parameter. In HD there are
several choices of spinning axis and there is a multitude of angular
momentum parameters, each referring to a particular spinning plane.
We concentrate on the simplest case for which there is only one
angular momentum parameter, that we shall denote by $a$. Next we
perform the similar complex coordinate  transformation, in the HD,
as used by Newman and Janis \cite{nja} by defining a new set of
coordinates $(u',r',\theta_i')$, where $i=1, \;\ldots, (N+1)/2$ by
the relations
\begin{figure*}

\begin{tabular}{|c|c|c|c|}
\hline
\includegraphics[width= 6 cm, height= 5 cm]{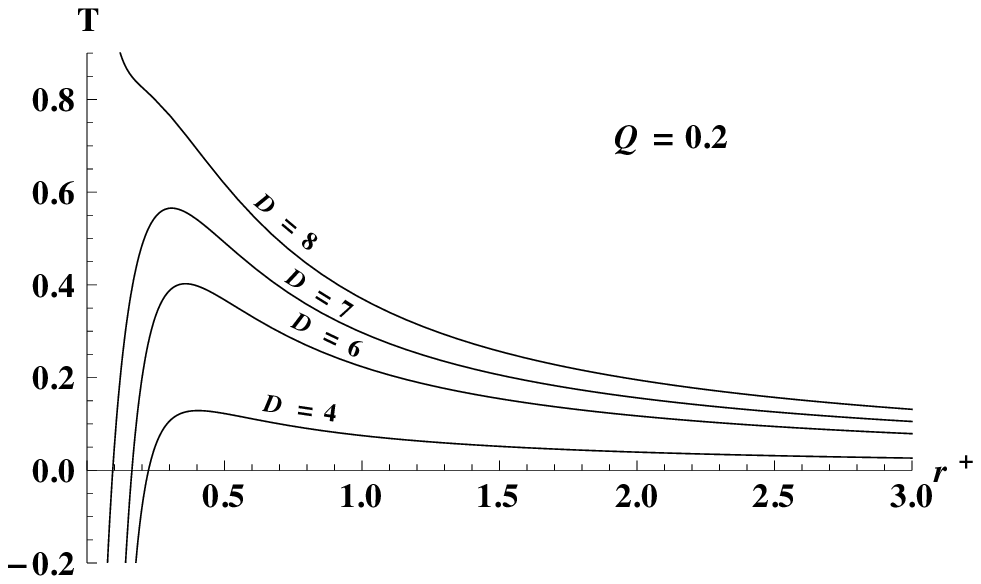}&% Here is how to import EPS art
\includegraphics[width= 6 cm, height= 5 cm]{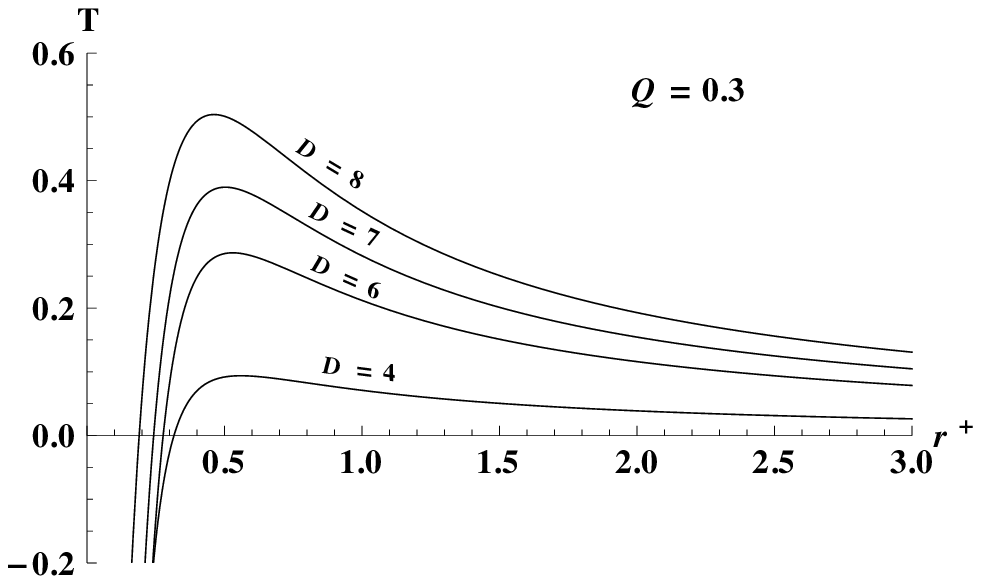}&
\includegraphics[width= 6 cm, height= 5 cm]{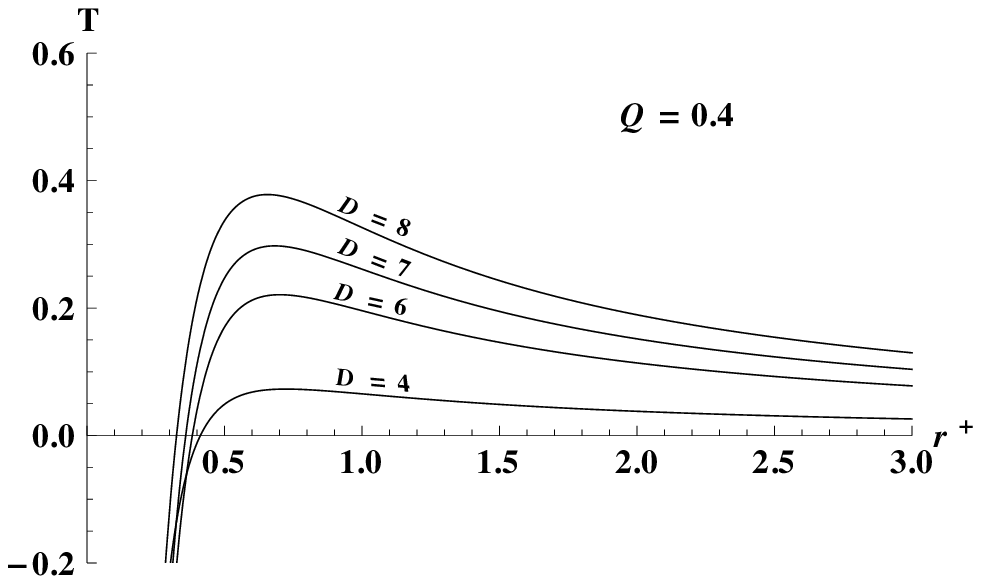}
\\
\hline
\end{tabular}
\caption{\label{Temperature}  The temperature profile shown as function of $r^+$ for different dimensions $D$ with three different values of YM gauge charge parameter $Q$.}
\end{figure*}

\begin{figure*}

\begin{tabular}{|c|c|c|c|}
\hline
\includegraphics[width= 6 cm, height= 5 cm]{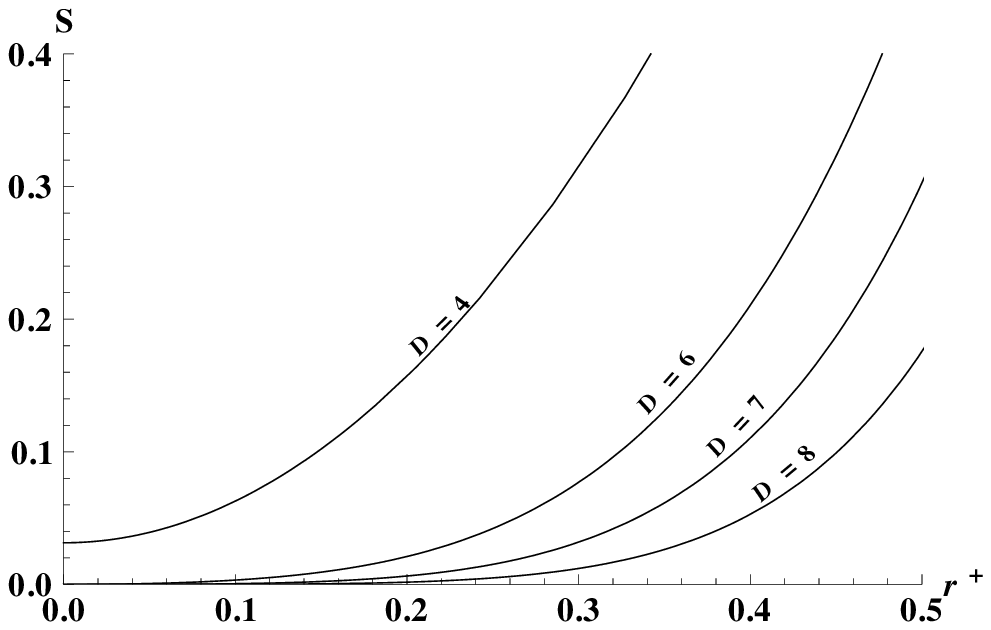}&% Here is how to import EPS art
\includegraphics[width= 6 cm, height= 5 cm]{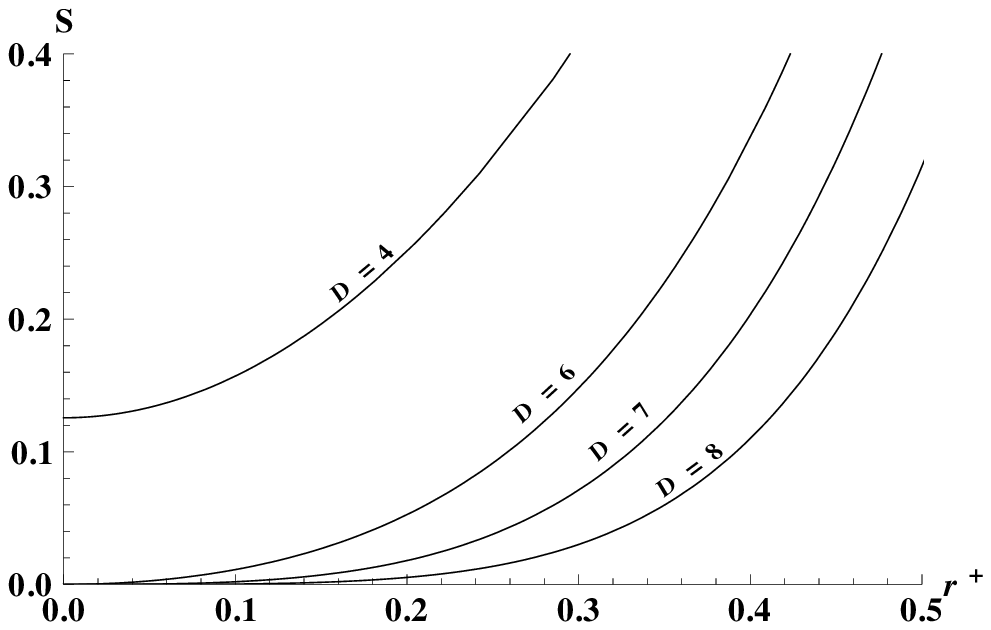}&
\includegraphics[width= 6 cm, height= 5 cm]{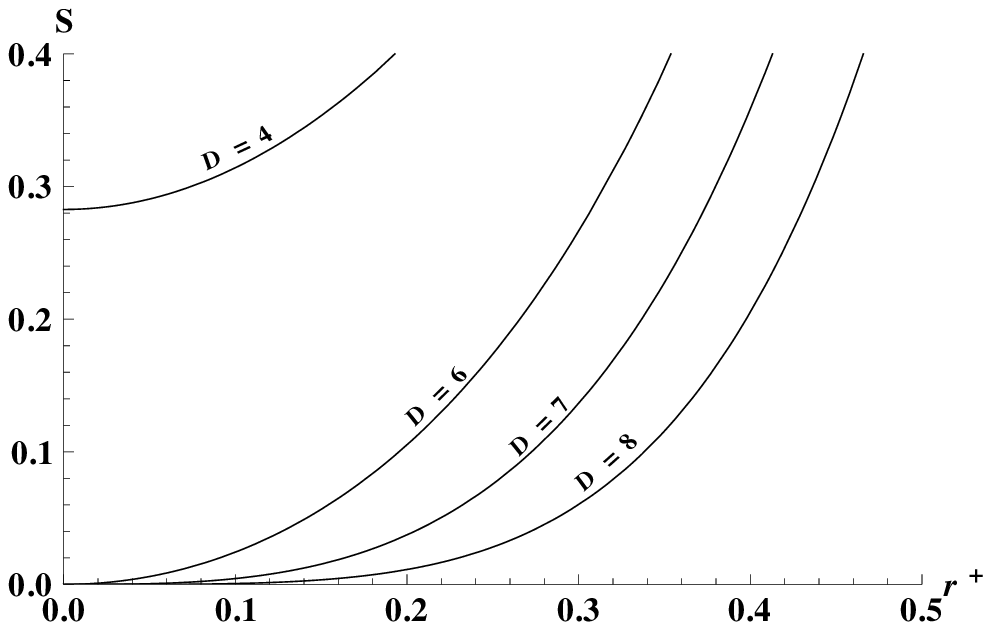}
\\
\hline
\end{tabular}
\caption{\label{entropy}  The behavior of entropy as a function of
horizon radius $r^+$ for different dimensions $D$ with three
different values of rotation parameter $a$.}
\end{figure*}
\begin{equation}\label{transf}
{x'}^{a} = x^{a} + ia (\delta_r^{a} - \delta_u^{a})
\cos\theta_1 \rightarrow \\ \left\{\begin{array}{ll}
u' = u - ia\cos\theta_1, \\
r' = r + ia\cos\theta_1, \\
\theta_i' = \theta_i \end{array}\right.
\end{equation}
Simultaneously let null vieltrad vectors $Z^a$ undergo a transformation $Z^a = Z'^a{\partial x'^a}/{\partial x^b} $ in the usual way, we obtain

\begin{eqnarray*}
%\nonumber to remove numbering (before each equation)
l^a &=& \delta^a_r, \\
n^a &=& \left[ \delta^a_u - \frac{1}{2} \left(1- \frac{\mu}{r^{N-2}\Sigma}- \frac{Q_0^2}{\Sigma}\right) \delta^a_r \right], \\
 m_{k}^a &=& \frac{1}{\sqrt{2}(r+ia\cos\theta_1)\sin\theta_{1}\sin\theta_{2},\; \ldots \;, \sin\theta_{(k-1)}} \nonumber  \\ &&
   \times \left(ia(\delta^a_u-\delta^a_r)\sin\theta_1 + \delta^a_{\theta_{k}} + \frac{i}{\sin\theta_{k}} \delta^a_{\theta_{(k+1)}} \right),\\
 \end{eqnarray*}
where we have dropped the primes. From the new null vieltrad, a new metric is discovered using (\ref{NPmetric}), which can be written as
\begin{eqnarray}
ds^2  &=&  \left(1-\frac{\mu}{r^{N-2}\Sigma}-
\frac{Q_0^2}{\Sigma}\right)du^2+2dudr- 2a\sin^2\theta_1 dr d\theta_2
\nonumber \\ &&  -\Sigma d\theta_1^2
-\left[(r^2+a^2)+\left(\frac{\mu}{r^{N-2}\Sigma}+\frac{Q_0^2}{\Sigma}\right)a^2
\sin^2\theta_1\right] \nonumber \\ &&  \times\sin^2\theta_1 d\theta_2^2
~-~2a\left(\frac{\mu}{r^{N-2}\Sigma} +\frac{Q_0^2}{\Sigma}\right)
\sin\theta_1^2 du d\theta_2 \nonumber \\ && - r^2 d\Omega_{N-1}^2,  \label{metric1}
\end{eqnarray}
where $\Sigma=r^2+a^2\cos \theta_1^2$. Thus, we have obtained
spinning BH corresponding to HD EYM BH. Also note that the derived HD metric~(\ref{metric1}) via
NJA is in Kerr like coordinates \cite{Xu}.  A
further simplification is made on taking coordinate transformation as in Ref. \cite{Xu}. This transformation leaves only one off-diagonal element
and we arrive at the following:
\begin{eqnarray}
ds^2  & = & \left(\frac{\Delta
-a^2\sin^2\theta_1}{\Sigma}\right)dt^2- \frac{\Sigma}{\Delta}dr^2  +
 2a  \nonumber \\ & &  \times \left[1-\left(\frac{\Delta
-a^2\sin^2\theta_1}{\Sigma}\right)\right]  dt d\theta_2 - \Sigma
d\theta_1^2 \nonumber \\
& &   - \left[\Sigma + a^2\sin^2\theta_1 \left(2-\frac{\Delta
-a^2\sin^2\theta_1}{\Sigma}\right)\right]   \nonumber \\
& &   \times \sin^2\theta_1 d\theta_2^2  - r^2 \cos^2\theta_1
d\Omega_{N-1}^2, \label{eq:mtc}
\end{eqnarray}
where on substituting back the value of $Q_0$ in terms of $Q$, $\Delta$ reads
\begin{eqnarray}
\Delta &=& r^2+a^2-\frac{\mu}{r^{N-2}}- \frac{N}{(N-2)}Q^2. \label{De}
\end{eqnarray}
Eq.~(\ref{eq:mtc}) is in Boyer-Lindquist coordinates. Here, we have
also introduced
\begin{eqnarray*}
\Delta = a^2 \sin^2\theta_1 + \Sigma G(r,\theta_1), \nonumber\\
G(r,\theta_1) = \frac{\Delta-a^2 \sin^2\theta_1}{\Sigma}.
\end{eqnarray*}
Thus, we are able to generate HD axisymmetric solution starting with
HD static spherically symmetric EYM BH solutions using the  approach
originally proposed by Newman-Janis \cite{nja}, i.e., we have an
explicit YM gauge charged HD spinning BH solution. One can see that
HD metric (\ref{eq:mtc})  as behavior of a metric produce from
spinning charged source. In the limit $N=1$, the geometry of
solution (\ref{eq:mtc}) is precisely of the Kerr-Newman form
\cite{Xu} and the charge that determines the geometry is YM gauge
charge. Thus we have exact HD Kerr-Newman like solution,  but $Q$
corresponds to magnetic charge. Hereafter, we refer the solution
(\ref{eq:mtc})  spinning HD EYM BH solutions. For vanishing YM gauge
charge $Q=0$, one recovers Myers-Perry BH solution discussed in
\cite{Myers}. The Reissner-Nordstr$\ddot{o}$m BH are recovered in
the limit $a=0$ and $N=1$. The HD EYM BH \cite{shmh081} are
discovered for the vanishing spinning parameter $a=0$.   It is nice
to see that the HD metric (\ref{eq:mtc}) gives all correct limit. We
have to still ensure that  HD metric (\ref{eq:mtc}) indeed solves
EYM equations (\ref{eq:ee}).   The NJA widely used in general
relativity and is correct for 4D. If $D \neq 4$, the trace of EMT
(\ref{emt}) tensor is not equal to zero, then $R\neq 0$.  This makes
problem rotating in HD-EYM case more complicated.  It is well known
that there is also dispute over the existence of Kerr-Newman in
higher dimensions \cite{Aliev}.  The Kerr-Newman solution  in
general  relativity is obtained  either using  the metric ansatz in
the Kerr-Schild form or applying the method of complex coordinate
transformation to a non-rotating charged black hole.  However, it
turns out that both procedure leads to same metric \cite{Aliev1}. We
have presented HD version of rotating EYM BH solution using the
complex transformation and we demonstrated that it has all
properties of a rotating BH. We have justified that the properties
of metric (15) are very similar to Kerr/Kerr-Newman. In particular,
we have also calculated event horizon and time like limit surface
and they are also similar to Myers-Perry BH

Eq. (\ref{eq:mtc}) has parameters $\mu$ and $a$ which are respectively related to mass $(M)$ and angular momentum $(J)$ via relations :-
\begin{eqnarray*}
M &=& \frac{(N+1)}{16\pi} A_{N+1}\mu, \;\; J = \frac{1}{8\pi}A_{N+1}\mu a,
\end{eqnarray*}and \begin{equation}
\frac{M}{J} = \frac{(N+1)}{2}a.
\end{equation}
The determinant $g$ of the metric (\ref{eq:mtc}) gives
\begin{eqnarray}
\sqrt{-g} = \sqrt{\gamma} \Sigma r^{N-1} \sin\theta \cos^{N-1}\theta,
\end{eqnarray}
and $A_{N+1}$ is the area of unit $(N+1)$ sphere which is given by
\begin{eqnarray}
A_{N+1}=\int_{0}^{2\pi}d\theta_2 \int_0^\pi \sin\theta_1
\cos^{N-1}\theta_1 d\theta_1  \\ \nonumber \times
\huge\prod_{i=3}^{N-1}\int_{0}^{\pi}\sin^{(N-1)-i}\theta_{i}d\theta_{i}
 = \frac{2\pi^{(N+2)/2}}{\Gamma(N+2)/2}.
\label{a}
\end{eqnarray}
The angular velocity at the horizon is given by
\begin{equation}
\Omega_H = \frac{a}{r^+{^2}+a^2}.
\end{equation}
Area of the event horizon (EH) for metric (\ref{eq:mtc}) can be given by the standard definition of the horizon area \cite{rab} as:
\begin{eqnarray*}
%\nonumber to remove numbering (before each equation)
&& A_H=\int_{\rho(r=r^+)}\sqrt{\eta} d\theta_1 d\theta_2 d^{N-1}\rho,
\end{eqnarray*}which trivially solves to \begin{eqnarray*}
&& A_H = r^+{^{N-1}}(r^+{^2} +a^2) A_{N+1}. \\
\end{eqnarray*}
The entropy of the BH typically satisfies the area  law of
the entropy which states  that the entropy of the BH is one
fourth of the area of EH \cite{Bek}. The horizon area and the surface gravity of the solution are related to the entropy and the temperature, respectively, $S=A_H/4$ and $T=\kappa/{2\pi}$. Thus the expression for entropy and temperature of the spinning HD EYM BH on the horizon are
\begin{eqnarray}
S &=& \frac{r^+{^{N-1}}(r^+{^2} + a^2) A_{N+1}}{4}, \nonumber \\
T &=& \frac{(N-2)(r^+{^2}+a^2)+2r^+{^2}-NQ^2}{4\pi r^+(r^+{^2}+a^2)}. \nonumber \end{eqnarray}In the appropriate limit the physical quantities derived above
reproduces the corresponding  qualities of Myers-Perry BH when
$Q=0$, of Kerr-Newman BH if $N=1$ and when both
$Q=0$ and $N=1$, we get these quantities associated with Kerr
BH. In Figs. (\ref{Temperature}) and (\ref{entropy}), we plot temperature and entropy of the spinning HD EYM BH respectively. It is interesting to note from Fig. (\ref{Temperature}) that at low value of YM gauge charge temperature is at maximum in HD but as $Q$ is increased temperature starts increasing from a minimum value and  on reaching at a maximum value, starts decaying. We note that the rate of decrease in temperature slows down with the increase in $Q$. The entropy for 4D case, $N=1$, is always positive even for vanishing horizon radius, $r^+=0$, but it is zero in HD case for $r^+=0$. The dependence of entropy on the horizon radius, $r^+$, is shown in Fig. (\ref{entropy}) which also confirms the area law for our solution, i.e., entropy is increasing with the radius of horizon.
\section{Horizon Properties}
It is natural to discuss not only spinning BH solutions but their
various properties. It is known that the structure of a spinning BH is much
different from that of a stationary BH. The EH of a spinning BH is smaller
than the EH of an otherwise identical but non-spinning one.
Similar to Kerr solutions in asymptotically flat spacetimes the
above metric has two types of
horizons like hypersurface: a stationary limit surface (SLS) and an
EH. Within the
stationary limit, no particles can remain at rest, even though they
are outside the EH. We shall explore the two horizons SLS
and EH of spinning HD EYM BH, and also
discuss the effects which comes from the YM gauge charge and also
due to spacetime dimensions.

Let us now address the horizon properties of the solution, beginning
with SLS. The SLS is the boundary of the region in which an observer
traveling a long time-like curve can follow the orbits of the
asymptotic time translation Killing vector $\partial/
\partial t$ and so remain stationary with respect to infinity. Physical observers cannot follow the orbit of  $\partial/\partial t$ beyond the EH surface since in that region they are spacelike orbits. On this surface
the Killing vector $\partial/\partial t$ is null. They are surfaces
of infinite redshift, and  for the spinning HD EYM BH requires that
prefactor $g_{tt}$ of the $dt^2$ term in metric vanish. It follows
that SLS will satisfy
\begin{equation}
 r^{N}+\left[a^2\cos^2\theta
-\frac{N}{(N-2)}Q^2\right] r^{N-2}-\mu=0. \label{eq:gtt}
\end{equation}

On the other hand, surfaces at which a particle traveling on a
timelike curve from a point on or inside the surface cannot get
outside the surface and so cannot get out to infinity is an EH. EH
is a solution of $\Delta=0$ and thus it must satisfy
\begin{equation}\label{eq:del}
r^{N}+\left[a^2-\frac{N}{(N-2)}Q^2\right] r^{N-2}-\mu=0.
\end{equation}
\begin{figure*}

\begin{tabular}{|c|c|c|c|}
\hline
\includegraphics[width= 6 cm, height= 5 cm]{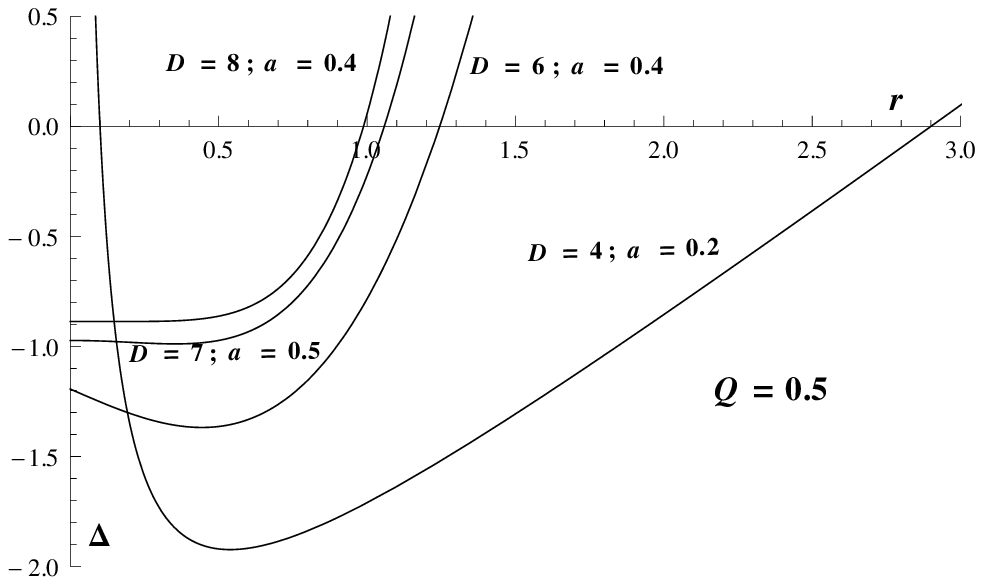}&% Here is how to import EPS art
\includegraphics[width= 6 cm, height= 5 cm]{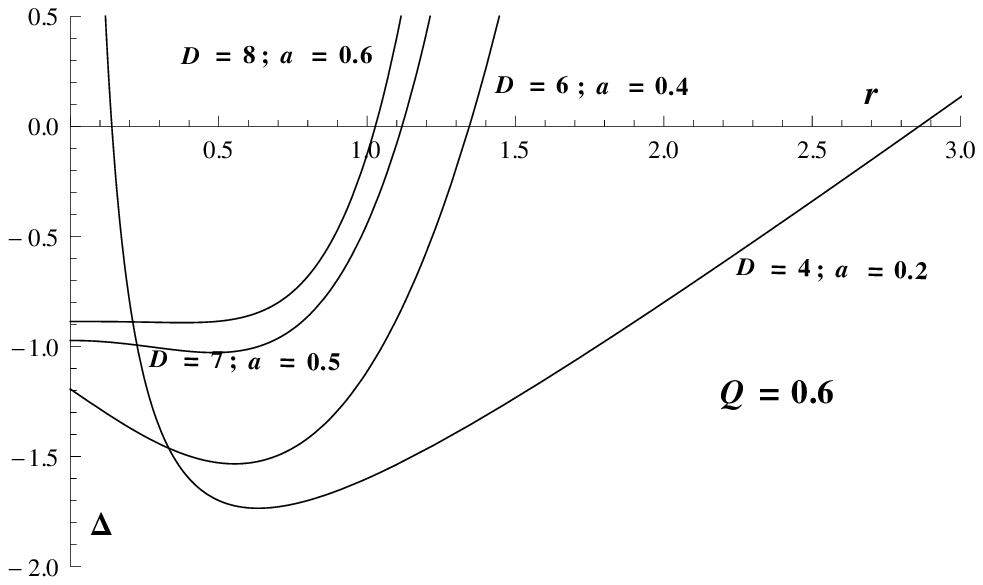}&
\includegraphics[width= 6 cm, height= 5 cm]{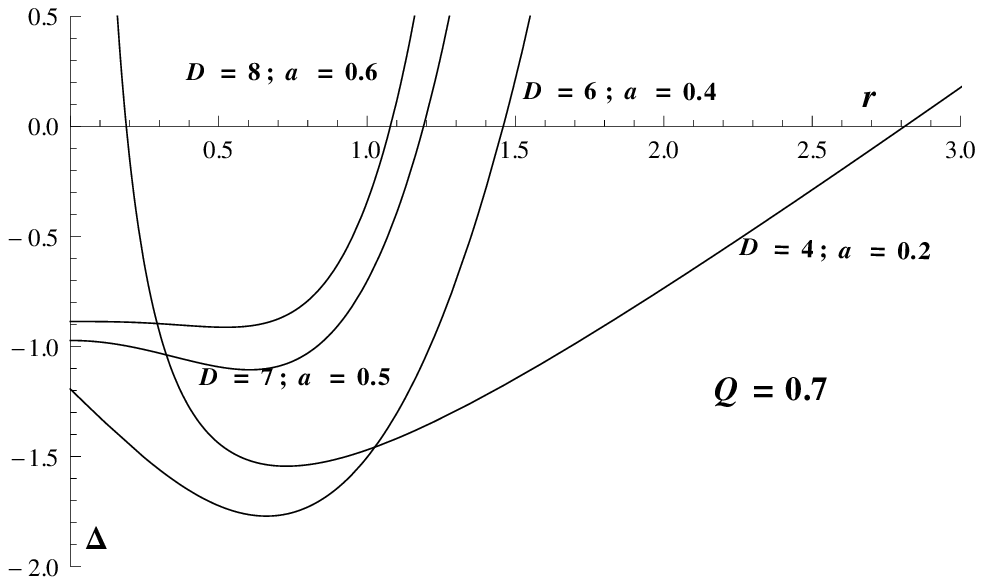}
\\
\hline
\end{tabular}

\caption{\label{EHa} Plot of $\Delta(r)$ to show the behavior of
horizon of spinning HD EYM BH for different dimensions $D$ with
three different values of YM gauge charge parameter $Q$. Here we
choose $M=1$.}
\end{figure*}
\subsubsection{4D Case}
When $N=1$, i.e., in 4D, recalling that $ \mu= 2M$, we recover the well- known results for the Kerr-Newman metric:
\begin{eqnarray}
ds^2 &=& \left(1-\frac{2M
r}{\Sigma}+\frac{Q^2}{\Sigma}\right)dt^2-\frac{\Sigma}{\Delta}dr^2-\Sigma
d\theta^2 \\  \nonumber && ~-~ \left[(r^2+a^2) + \left(\frac{2M
r}{\Sigma}-\frac{Q^2}{\Sigma}\right)a^2 \sin^2\theta\right]
\sin^2\theta d\phi^2 \\ \nonumber && ~+~ 2a\left(\frac{2M
r}{\Sigma}-\frac{Q^2}{\Sigma}\right)  \sin^2\theta dt d\phi,
\label{met1}
\end{eqnarray}
with $\theta_1=\theta$, $\theta_2=\phi$ and in (\ref{eq:mtc}) $\mu=2M$. The Eq. (\ref{eq:del}), for $N=1$, is simplified to \begin{equation}
\Delta =  r^2+a^2 - 2Mr + Q^2,
\end{equation} which admit solution $r^{\pm}$, identified as outer and inner EH. The EH of Eq.~(\ref{met1}) are
\begin{equation} \label{EH}
r^{\pm}{_{EH}}=M \pm \sqrt{(M^2-Q^2)-a^2}.
\end{equation}
If $a^2<(M^2-Q^2)$, there exist two horizons, when
$a^2\rightarrow(M^2-Q^2)$,  two horizons coincides, i.e., the extremal case and if
$a^2>(M^2-Q^2)$, then there exists no horizon, i.e., one has naked singularity. For SLS, $N=1$ in Eq. (\ref{eq:gtt}), it reduces to
\begin{equation}\label{gr}
{r}^{2}-2\,M r+{a}^{2} \cos ^{2}\theta  + Q^{2}  =0,
\end{equation}
 which trivially solves to
\begin{eqnarray} \label{SLSkn}
% \nonumber to remove numbering (before each eQuation)
r^{\pm}{_{SLS}} &=& M \pm \sqrt {(M^{2} -Q^2)
 - {a}^{2} \cos^{2} \theta}.
 \nonumber \\
\end{eqnarray}
These are regular outer and inner SLSs for a Kerr-Newman
BH when $a^2\cos^2\theta<(M^2-Q^2)$, and further  in the non-spinning limit $a
\rightarrow 0$, both SLS and EH coincides to
\begin{eqnarray} \label{SLSkn1}
% \nonumber to remove numbering (before each eQuation)
r^{\pm} &=& M \pm \sqrt {  M^{2}  -
   Q^{2}}, \nonumber \\
\end{eqnarray}
which are outer and inner EH of Reissner-Nordstr$\ddot{o}$m BH.  Thus the Kerr-Newman BH, in
the limit $a \rightarrow 0$, degenerates to Reissner-Nordstr$\ddot{o}$m BH.
\subsubsection{6D Case}  Eq. (\ref{eq:del}) for $N=3$ or $6D$ case reduces to
\begin{equation}
r^3+(a^2-3Q^2)r -\mu=0,
\end{equation}
which gives EH as
\begin{eqnarray*}
r^{+}{_{EH}} &=& \frac{(27\mu+\sqrt{729 \mu^2 + \delta})^{1/3}}{3\times 2^{1/3}}- \frac{2^{1/3}(a^2-3Q^2)}{(27\mu+\sqrt{729 \mu^2 +\delta})^{1/3}},
\end{eqnarray*}
with $\delta=4(3a^2-9Q^2)^3$.

Eq. (\ref{eq:gtt}) reduces to
\begin{equation}
r^3+(a^2\cos^{2}\theta- 3Q^2)r-\mu =0,
\end{equation} which can be solved to
\begin{eqnarray*}
r^{+}{_{SLS}} &=& \frac{(27\mu+\sqrt{729 \mu^2 +\delta})^{1/3}}{3\times 2^{1/3}}- \frac{2^{1/3}(a^2\cos^2 \theta -3Q^2)}{(27\mu+\sqrt{729 \mu^2 +\delta})^{1/3}},
\end{eqnarray*} with $\delta =4(3a^2 \cos^2\theta -9Q^2)$.
\subsubsection{7D Case}
Eq. (\ref{eq:del}), for 7D, reduces to
\begin{equation}
r^4+(a^2-2Q^2)r^2-\mu=0.
\end{equation}
So, we get the EH as
\begin{eqnarray*}
r^{+}{_{EH}} &=&  \sqrt{\frac{1}{2}\sqrt{4\mu+\delta_1}+\frac{1}{2}(2Q^2-a^2)},
\end{eqnarray*}
with $\delta_1=a^4+4Q^2-4a^2Q^2$.

Eq. (\ref{eq:gtt}) can be written as
\begin{equation}
r^4+(a^2\cos^{2}\theta- 2Q^2)r^2-\mu =0,
\end{equation}
which admits solution
\begin{eqnarray*}
r^{+}{_{SLS}} &=& \sqrt{\frac{1}{2}\sqrt{4\mu+\delta_2}+\frac{1}{2}(2Q^2-a^2\cos^2\theta) },
\end{eqnarray*}
with $\delta_2=a^4\cos^4\theta-4a^2\cos^2\theta+4Q^2$.
\begin{figure*}

\begin{tabular}{|c|c|c|c|}
\hline
\includegraphics[width=0.30\textwidth]{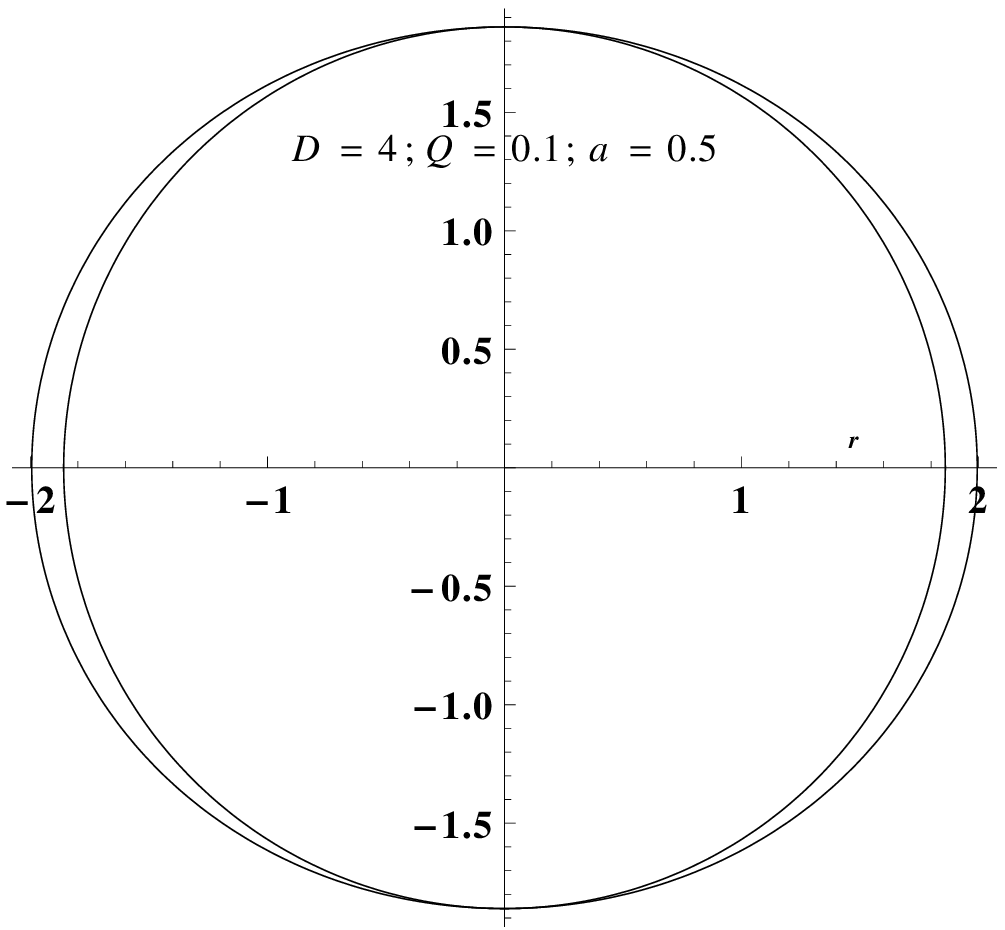}&% Here is how to import EPS art
\includegraphics[width=0.30\textwidth]{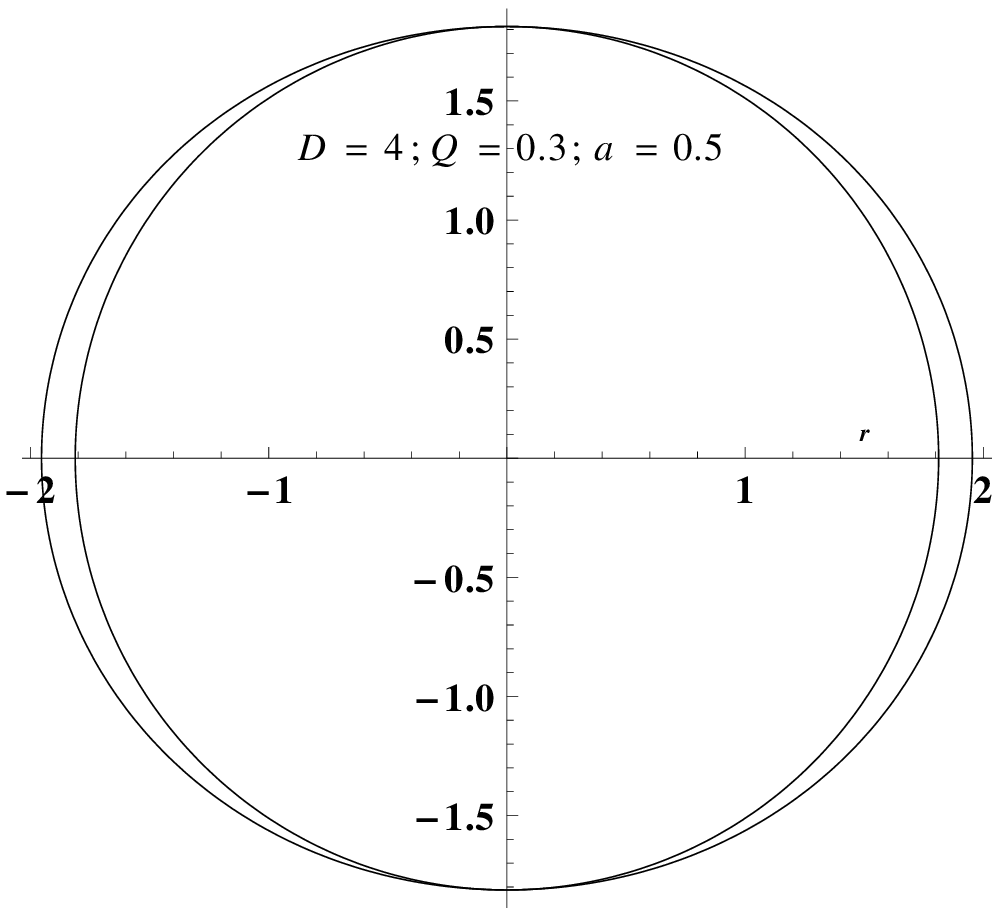}&
\includegraphics[width=0.30\textwidth]{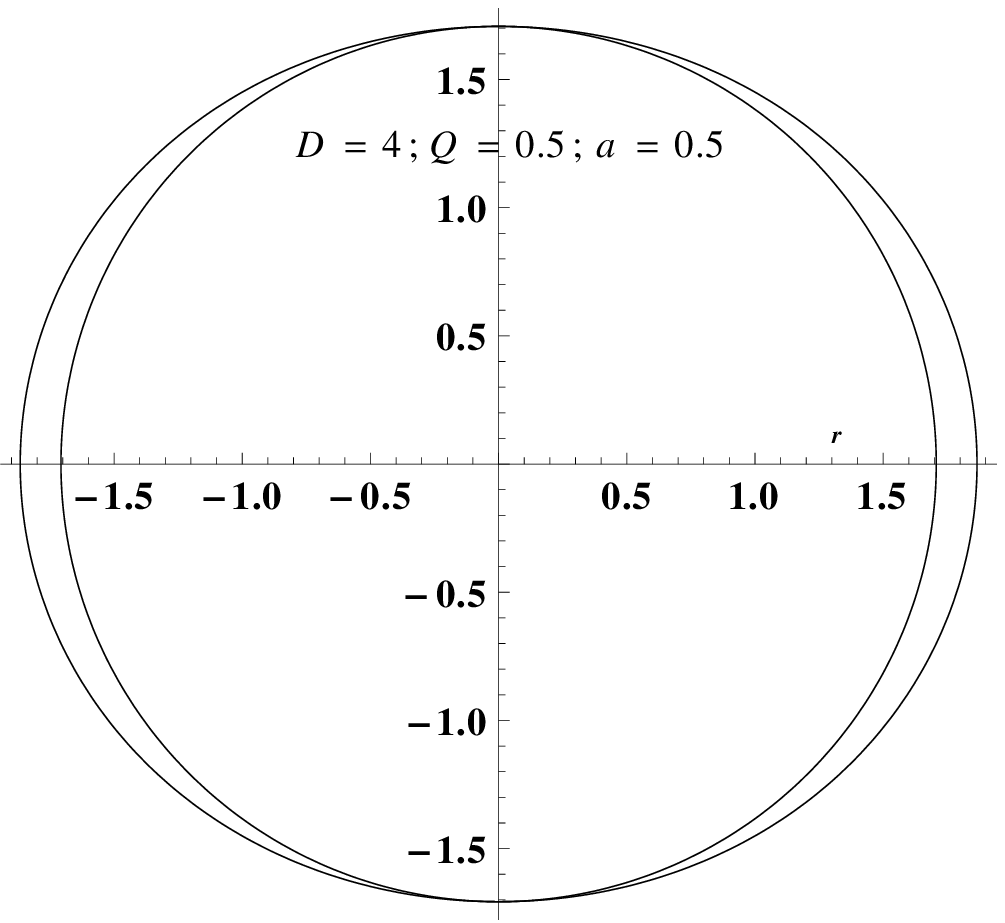}
\\
\hline
\includegraphics[width=0.30\textwidth]{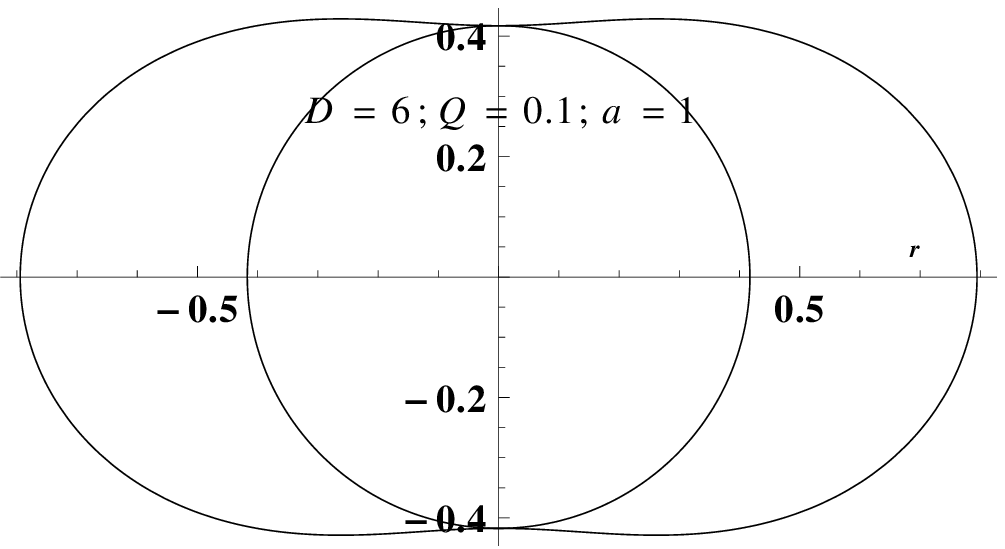}&% Here is how to import EPS art
\includegraphics[width=0.30\textwidth]{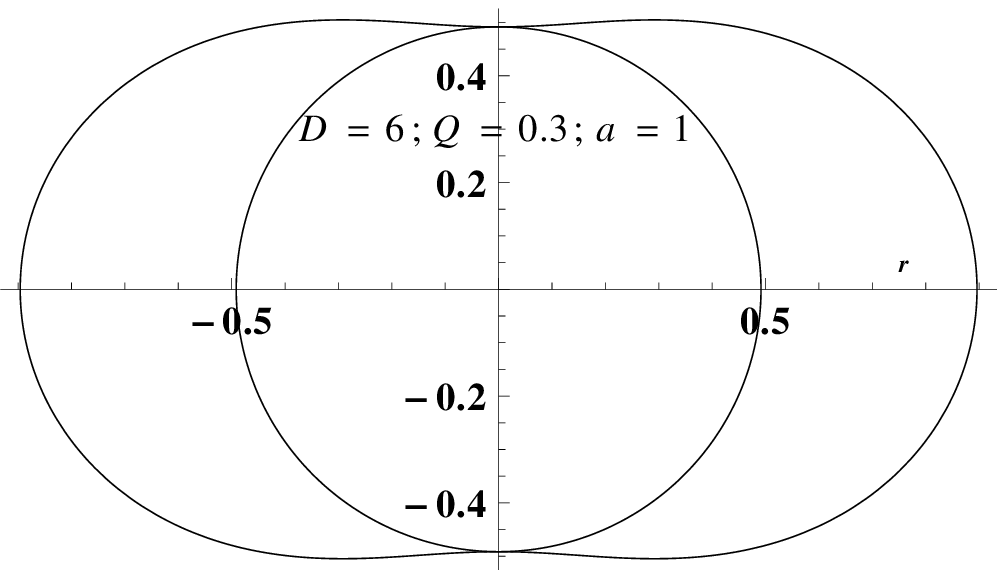}&
\includegraphics[width=0.30\textwidth]{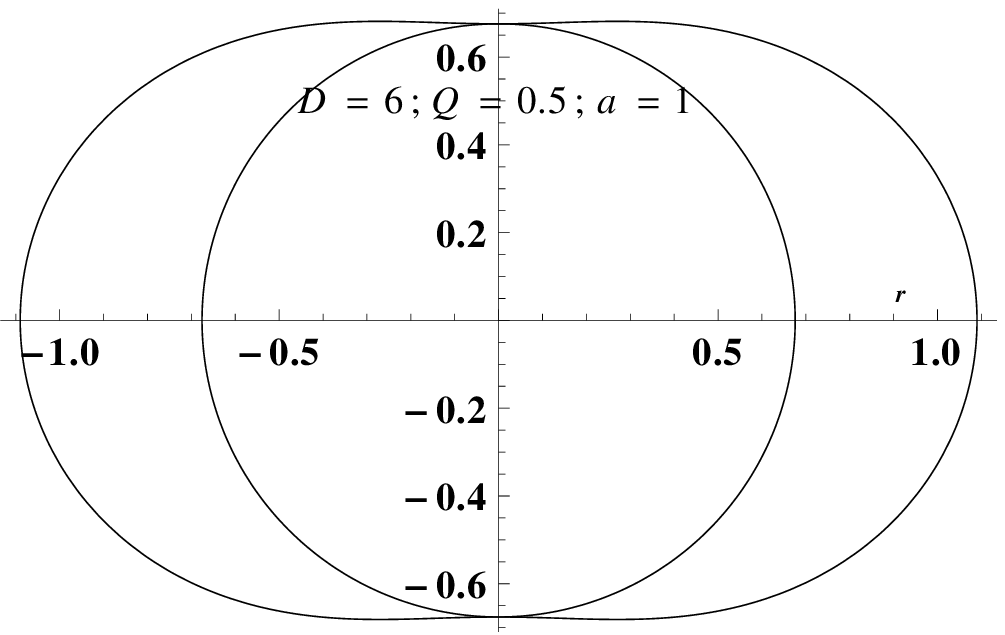}
\\
\hline
\includegraphics[width=0.30\textwidth]{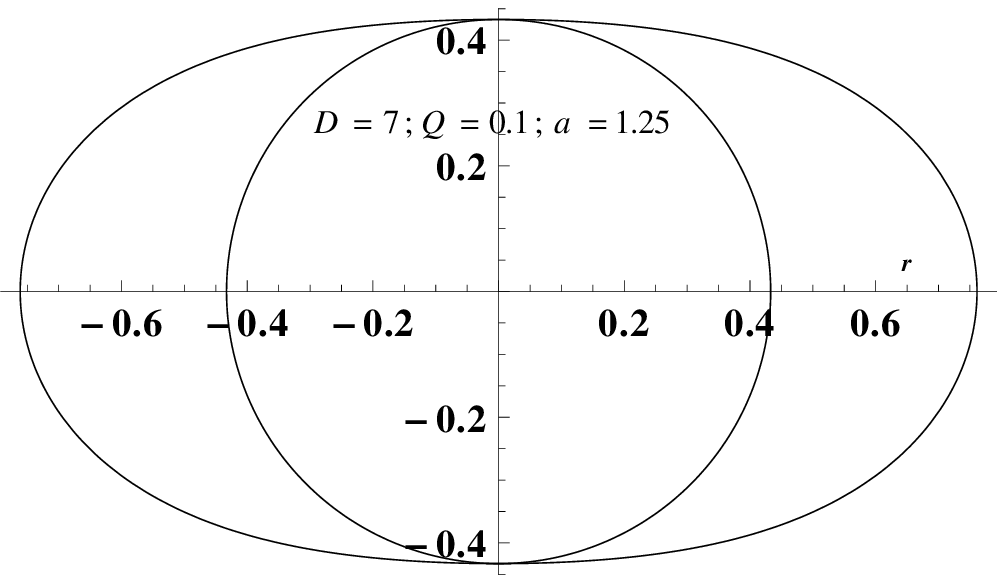}&% Here is how to import EPS art
\includegraphics[width=0.30\textwidth]{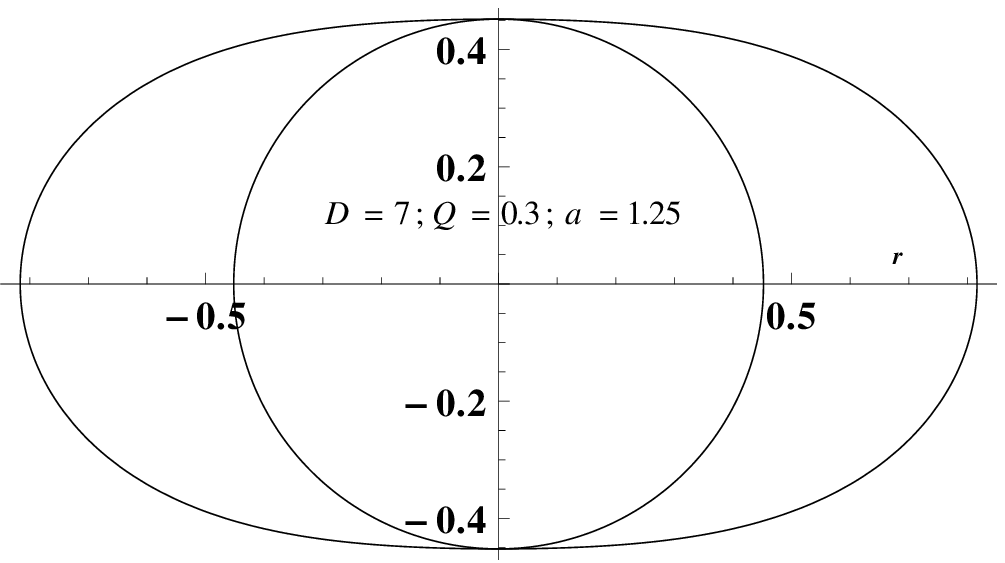}&
\includegraphics[width=0.30\textwidth]{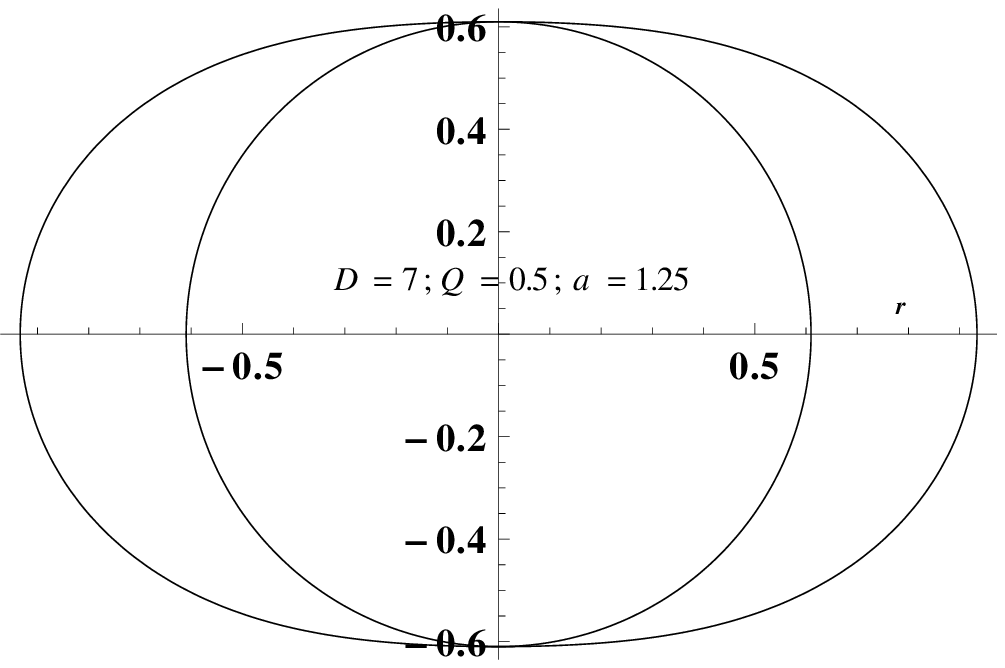}
\\
\hline
\end{tabular}
\caption{\label{ergo1}  Plots showing cross section of SLS and EH
and the variation of the ergosphere for different dimensions $D$ with three different values of YM gauge charge parameter $Q$.}
\end{figure*}

\begin{figure*}

\begin{tabular}{|c|c|c|c|}
\hline
\includegraphics[width=0.30\textwidth]{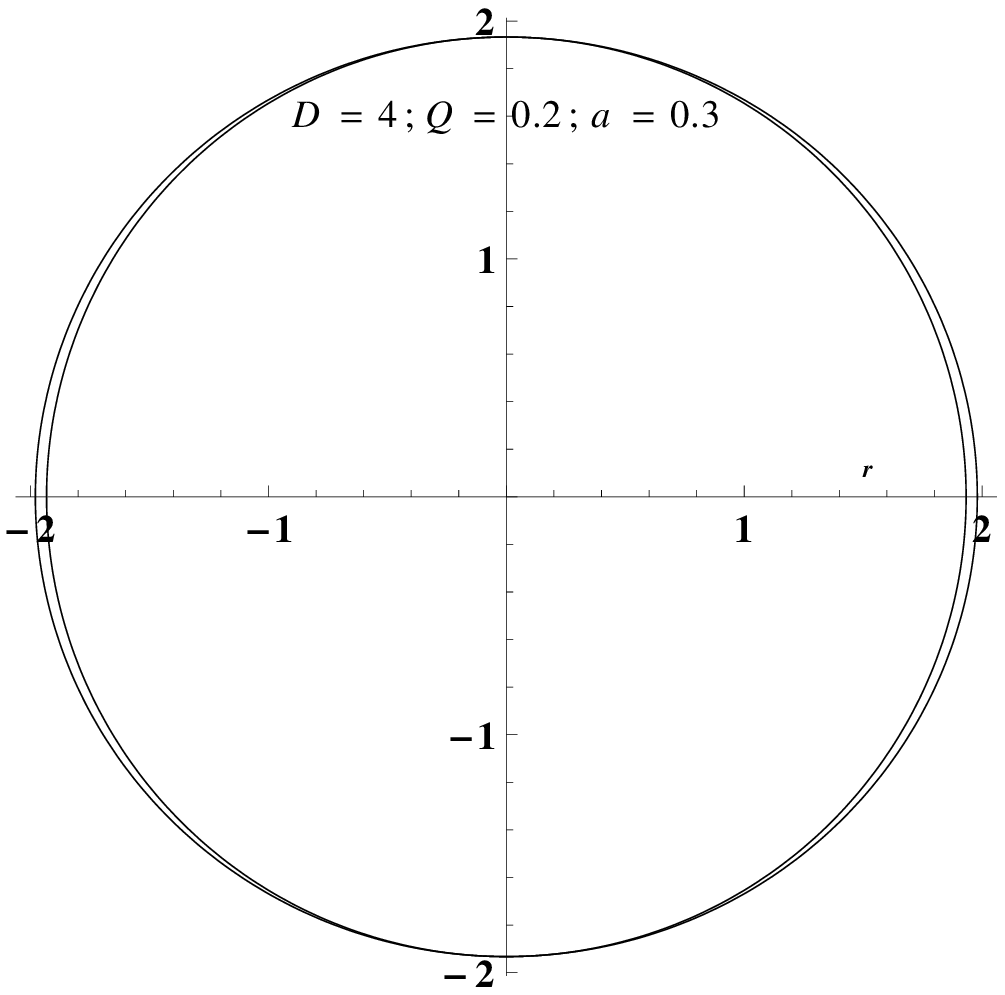}&% Here is how to import EPS art
\includegraphics[width=0.30\textwidth]{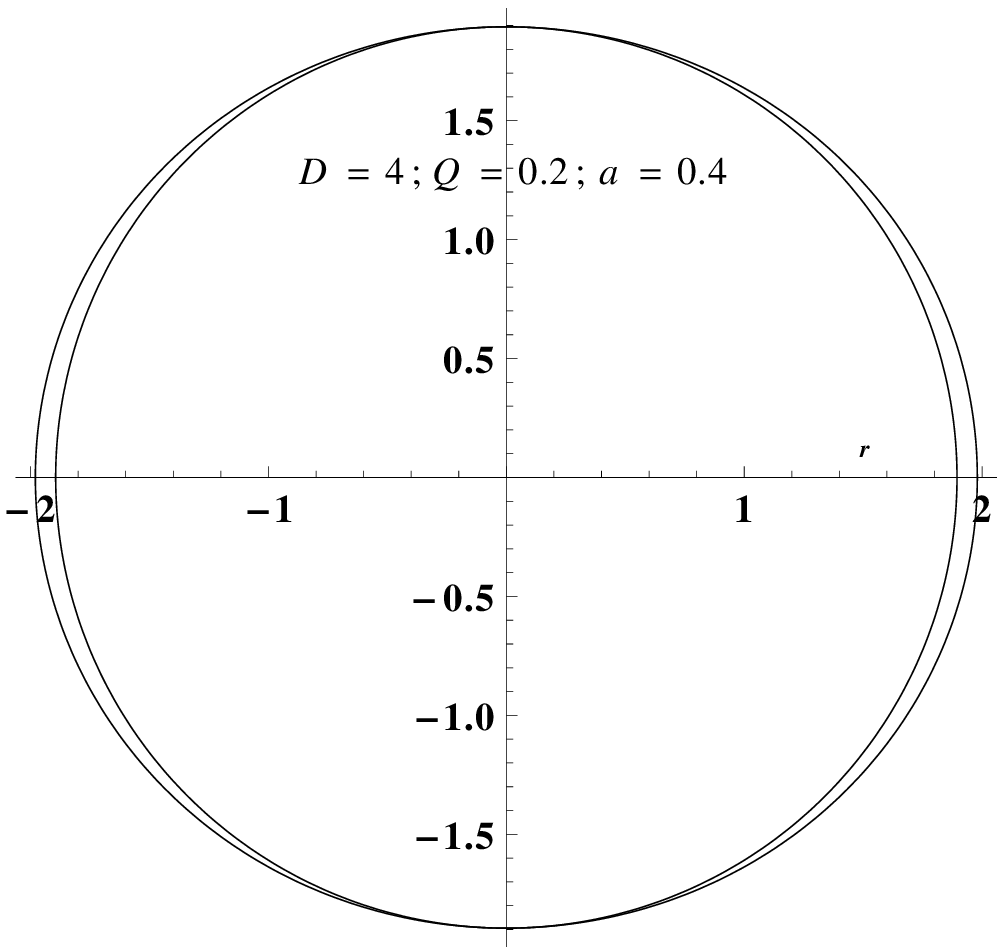}&
\includegraphics[width=0.30\textwidth]{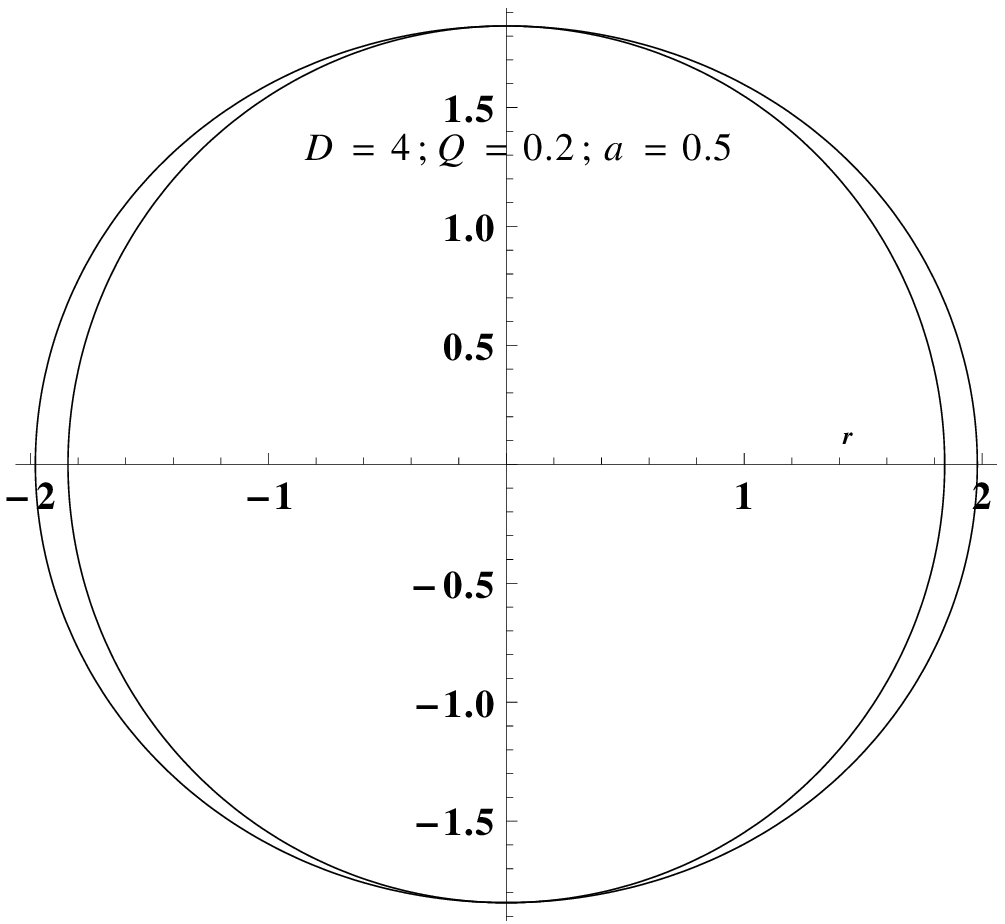}
\\
\hline
\includegraphics[width=0.30\textwidth]{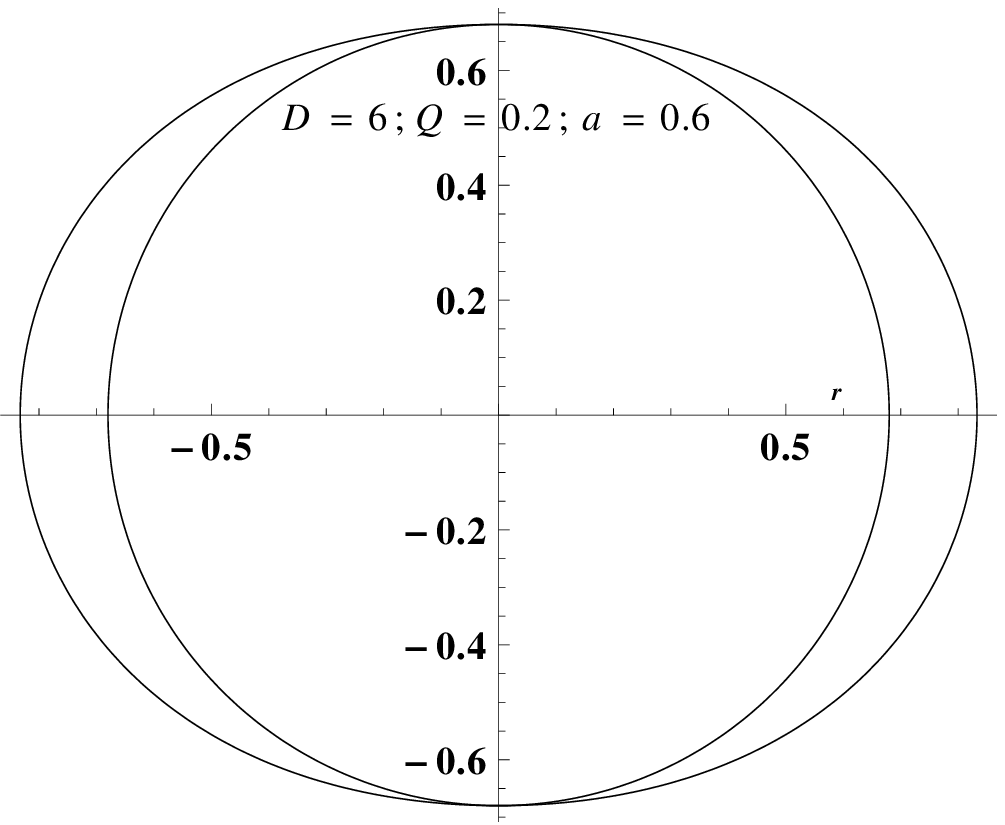}&% Here is how to import EPS art
\includegraphics[width=0.30\textwidth]{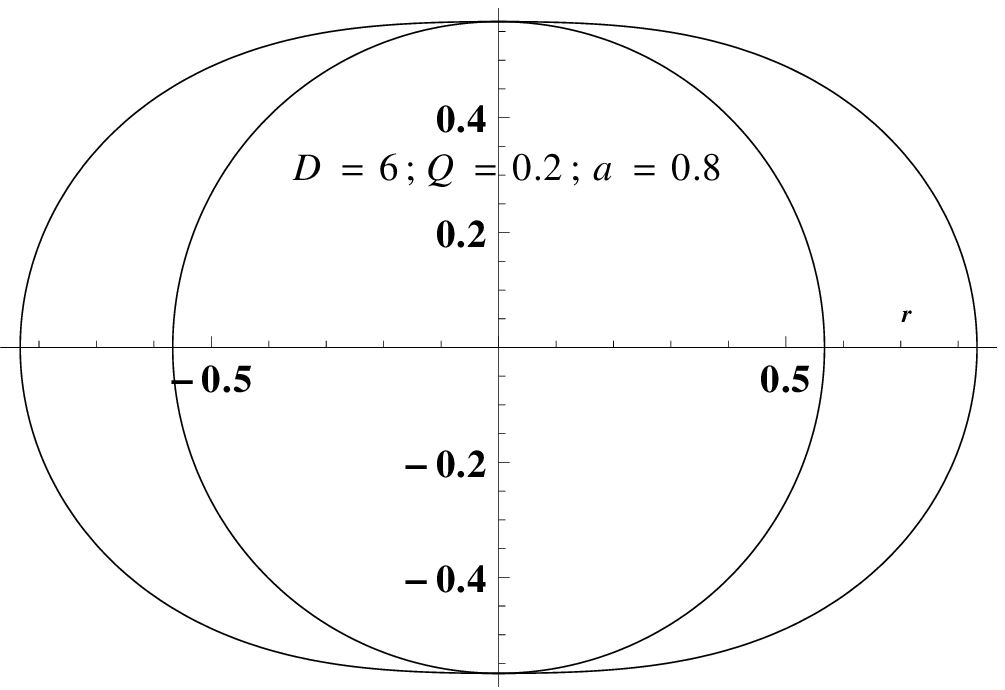}&
\includegraphics[width=0.30\textwidth]{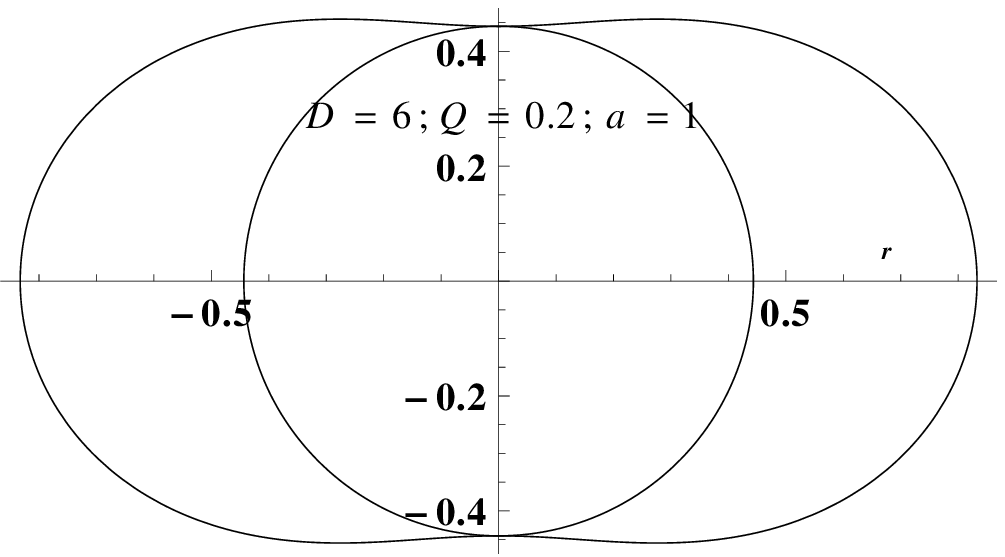}
\\
\hline
\includegraphics[width=0.30\textwidth]{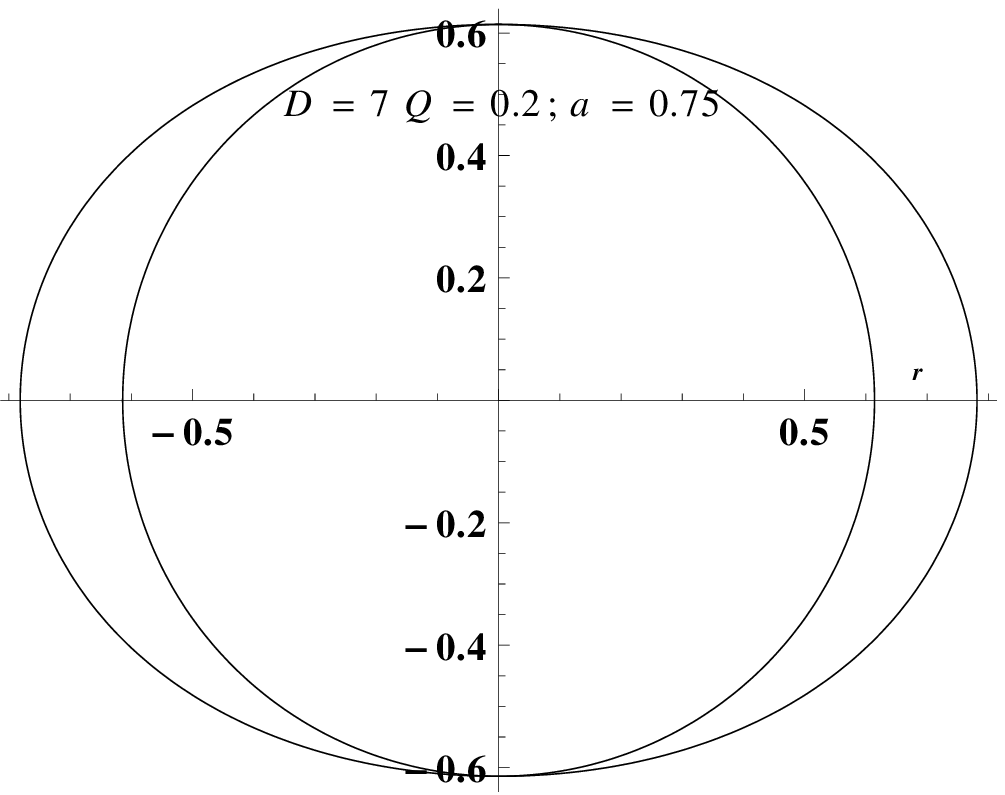}&% Here is how to import EPS art
\includegraphics[width=0.30\textwidth]{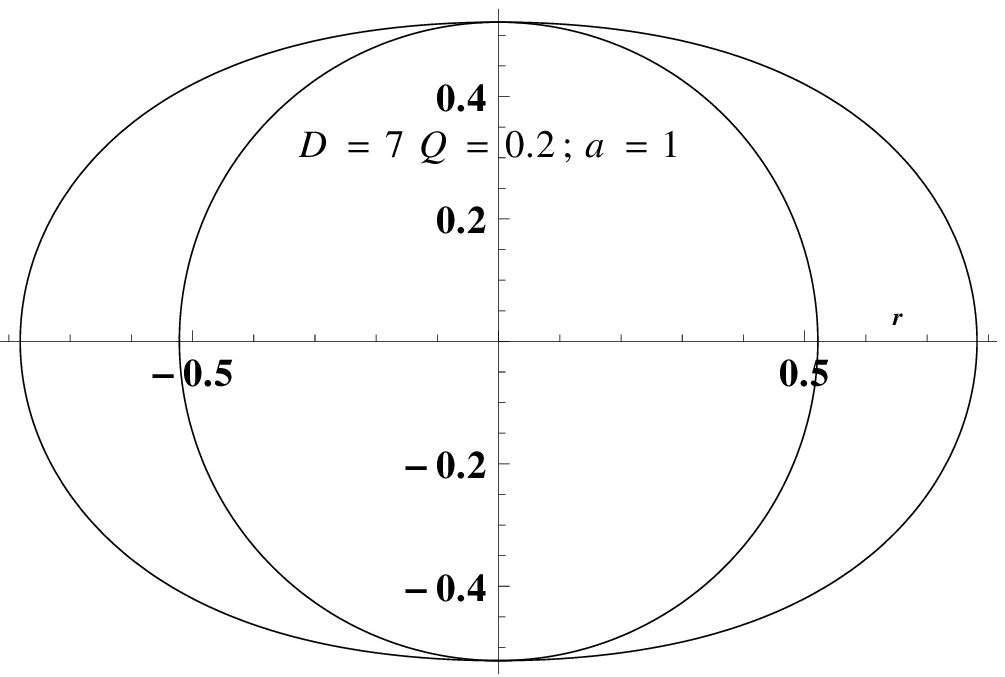}&
\includegraphics[width=0.30\textwidth]{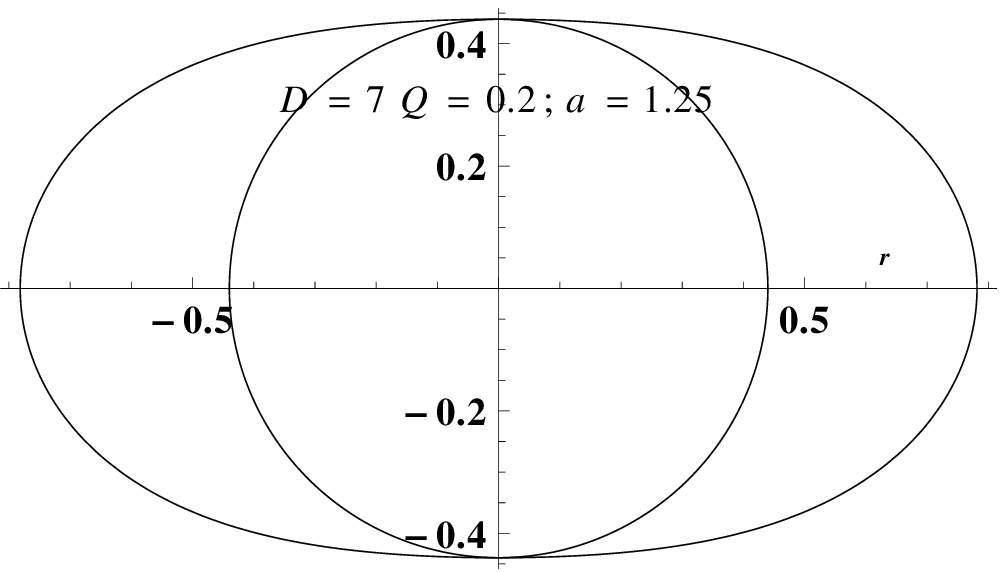}
\\
\hline
\end{tabular}
\caption{\label{ergo2}Plots showing cross section of SLS and EH and the dependence of the shape of ergosphere for different dimensions $D$ with three different values of rotation parameter $a$.}
\end{figure*}

Thus, the SLS and EH depends on the spacetime dimension. For HD we note that
\begin{equation}
\lim_{r\rightarrow 0} \Delta(r)= -\mu < 0,\;\;
\lim_{r\rightarrow +\infty} \Delta(r)= \infty,\;\; \Delta^{\prime}(r)\geq0.
\end{equation}
It is seen that $\Delta^{\prime}(r)\geq0$ for $D\geq6$. However it is seen that Eqs. (\ref{eq:gtt}) and (\ref{eq:del}) has just one positive root for HD $(D\geq6)$, i.e., just one EH and SLS in HD. This means that there is no extremal spinning BH when $D\geq6$. The effect of YM gauge charge on horizon is shown in Fig. ~(\ref{EHa}). It turns out that the radius of EH is decreasing with spacetime dimension. On the other hand, it increases with the value of YM gauge charge. One gets same result for SLS and we do not present them here.
\subsection{Ergosphere} For the Schwarzschild and Reissner-Nordstr$\ddot{o}$m
BH, it is possible that a traveler can approach arbitrarily close to
the EH whilst remaining stationary with respect to infinity. This is
not the case for the Kerr/Kerr-Newman BH. The spinning BH drags the
surrounding region of spacetime causing the traveler to spin
regardless of any arbitrarily large thrust that he can provide. The
ergoregion is the region in which this happens and is bounded by the
ergosphere.  On plotting the SLS and EH for spinning HD EYM BH,  it
can be verified that the SLS always lies outside the EH for all
dimensions $D$. The ergosphere is plotted in Figs. (\ref{ergo1}) and
(\ref{ergo2}). The ergosphere is defined to be the place where the
vectors $\partial t$ parallel to the \emph{time-axis} are not
timelike but spacelike. An ergosphere, thus, is a region of
spacetime where no observer can remain still with respect to the
coordinate system in question. Thus the ergosphere of a spinning HD
EYM BH, as in Kerr/Kerr-Newman BH, is bounded by the EH on the
inside and an oblate spheroid SLS, which coincides with the EH at
the poles and is noticeably wider around the equator.   It is the
region of spacetime where timelike geodesics remain stationary and
timelike particles can have negative energy relative to infinity. It
is theoretically possible to extract energy and matter from the BH
from ergosphere \cite{penrose}. In Fig. (\ref{ergo1}) we have shown
the dependence of the shape of the ergosphere on the YM gauge
charge. It is noticed that for $D=4$ the shape of the ergosphere is
increasing with the increase in YM gauge charge while decreasing for
$D\geq6$. This shows that shape of ergosphere is sensitive to the YM
gauge charge. In Fig. (\ref{ergo2}) the variation of the shape of
ergosphere with $a$ is shown. We note that the relative shape of the
ergosphere becomes more prolate thereby increasing area of the
ergosphere with rotation parameter $a$, i.e., the faster the BH
rotates, the more the ergosphere grows. This may have direct
consequence on the Penrose's energy extraction process
\cite{penrose}.
\section{Conclusion} In this paper, we have extended NJA in order to construct
 spinning BH from HD EYM BH. The method does not uses field equation but works on the HD spherical solution to generate spinning solutions. The algorithm is very useful since it directly allows us to construct spinning BH in HD, which otherwise could be extremely cumbersome. Originally, the NJA was applied to Reissner-Nordstr$\ddot{o}$m solution to obtain Kerr-Newman solution \cite{my}. The metric (\ref{eq:mtc}) is stationary, axisymmetric and asymptotically flat.  It depends on mass, YM gauge charge and spinning parameter which reduces to Kerr BH \cite{kerr} $(N=1, Q=0)$ and Myers Perry BH \cite{Myers} $(Q=0)$. The solution in 4D $(N=1)$ has precisely the geometry of Kerr-Newman \cite{my}, but the charge that determines the geometry is YM gauge charge. Also, it is easy to check that the metric (\ref{eq:mtc}) in 4D $(N=1)$ is solution of EYM Eq. (\ref{eq:ee}). Thus, we can say that a spinning BH solution of Einstein Maxwell equations is also solution of EYM, but the charge $Q$ is the YM gauge charge and not electric charge. However, this is not true in HD. Our spinning HD EYM BH solution deviate from HD Kerr-Newman \cite{Aliev} because $\Delta(r)$ for the latter is given by\begin{equation}
 \Delta_{KN}(r) =  r^2+a^2 - \frac{\mu}{r^{N-2}} + \frac{Q^2}{r^{2(N-1)}}, \label{KN}
\end{equation}
The corresponding $\Delta(r)$ in spinning HD EYM BH is given by Eq. (\ref{De}). The difference in the last term of (\ref{De}) and (\ref{KN}) is because of the fact that the charge term $Q^2/r^2$ in the solution (\ref{eq:me2}) is dimension independent, while it would go as $Q^2/r^{2(N-1)}$ in HD Reissner-Nordstr$\ddot{o}$m. It may be noted HD Reissner-Nordstr$\ddot{o}$m can be used to obtain HD Kerr-Newman using NJA \cite{Xu}. The two kinds of horizon like surfaces viz. SLS and EH are studied. In 4D, it turns out that there exist two horizon like surfaces corresponding to two positive roots which are identified as inner and outer horizons. However, in HD there exist only one positive root and thereby only one SLS and EH. It is interesting to see that the structure of SLS, EH and ergosphere are sensitive to YM gauge charge parameter $Q$.

The physical properties of the solutions have not yet been fully investigated; this being very severe job.  However, we are currently on this project.  We have also shown that the presence of YM gauge charge decreases the temperature with increase in gauge charge parameter Q. Such a change could have a significant effect in the thermodynamics of a BH. Hence, It will be of interest to see how YM gauge charge affects the thermodynamics by deriving Smarr-like relation and the first law, and also stability analysis. Further analysis of these solutions and the role of YM gauge charge and spacetime dimension in energy extraction process remains interesting issue to explore in the future.

\acknowledgements  Authors would like to thank
University Grant Commission (UGC) for major research project grant NO. F-39-459/2010(SR) and to IUCAA, Pune for kind hospitality while part of this work was being done.

\end{document}